\documentstyle[preprint,revtex]{aps}
\begin{document}
\draft
\begin{title}Isospin Dependence of the Spin Orbit Force
and Effective Nuclear Potentials
\end{title}
\author{M.M. SHARMA$^1$, G. LALAZISSIS$^2$, J. K\"ONIG$^2$, and P. RING$^2$}
\begin{instit}
$^1$Max Planck Institut f\"ur Astrophysik,
Karl-Schwarzschildstrasse 1, D-85740 Garching, Germany
\end{instit}
\begin{instit}
$^2$Physik-Department der Technischen Universit\"at M\"unchen,
D-85747 Garching, Germany
\end{instit}
\receipt{16 October, 1993}
\begin{abstract}
The isospin dependence of the spin-orbit potential is investigated
for an effective Skyrme-like energy functional suitable for
density dependent Hartree-Fock calculations. The magnitude of
the isopin dependence is obtained from a fit to experimental
data on finite spherical nuclei. It is found to be
close to that of relativistic Hartree models. Consequently,
the anomalous kink in the isotope shifts of Pb nuclei is well
reproduced.
\end{abstract}
\vskip 1 cm

The Hartree-Fock approach based upon phenomenological
density dependent forces\cite{VB.72,FQ.78,DG.80} has proved
to be very successful in the microscopic description
of ground state properties of nuclear matter and of finite
nuclei over the entire periodic table. In all these calculations
the spin-orbit potential has been assumed to be isospin
independent. Its only parameter, the strength, is usually adjusted to
the experimental spin-orbit splitting in spherical nuclei like
$^{16}$O. The exchange term, however, causes for nuclei with neutron
excess a strong isospin dependence of the correponding single-particle
spin-orbit field.

In recent years Relativistic Mean Field (RMF) theory\cite{SW.86} with
nonlinear self-interactions between the mesons has gained considerable
interest for the investigations of low-energy phenomena in
nuclear structure. With only a few phenomenological parameters
such theories are able to give a quantitative description of ground
state properties of spherical and deformed nuclei\cite{GRT.90,SNR.93}
at and away from the stability line. In addition, excellent agreement
with experimental data has been found recently also for collective
excitations such as giant resonances\cite{VBR.94} and for twin bands
in rotating superdeformed nuclei\cite{KR.93}. In many respects the
relativistic mean-field theory is regarded as similar to the
density-dependent Hartree-Fock theory of the Skyrme
type\cite{Thi.86,Rei.89}

Recently, however, detailed investigations of high precision data on
nuclear charge radii in Pb isotopes\cite{TBF.93,SLR.93} and of shell
effects at the neutron drip line\cite{Dob.94,SLH.94} have shown
considerable differences between the Skyrme approach and the relativistic
mean field theory. In fact, density dependent Hartree-Fock calculations
with Skyrme\cite{TBF.93} or Gogny forces\cite{Egi.94}, which used
so far antisymmetrized, isospin independent spin-orbit interactions,
were not able to reproduce the kink in the isotope shifts of Pb nuclei
(see Fig. \ref{F1}). On the other hand, this
kink is obtained in the RMF theory without any new adjustment of
parameters\cite{SLR.93}. Another considerable difference has
been found in theoretical investigation of shell effects in
very exotic Zr-isotopes near the neutron drip line: in conventional
non-relativistic Skyrme calculations
the shell gap at isotope $^{122}$Zr with the magic neutron configuration
N = 82 is totally smeared out\cite{Dob.94}, whereas relativistic
calculations using various parameter sets show  at N=82 a clear kink in
the binding energy as a function of the neutron number\cite{SLH.94}.
This difference is caused by the different spin-orbit splitting of
the single particle levels in these nuclei. Mass calculations within
the Finite Range Droplet Model (FRDM) \cite{MNM.94}, which are based on
an isospin dependent spin-orbit term carefully adjusted to experimental
data, are in excellent agreement with the relativistic predictions.

This gives us a hint, that one does not really need full relativistic
calculations in order to understand these differences between
conventional density dependent Hartree-Fock calculations and
RMF theory. A spin-orbit term with a properly chosen isospin
dependence might represent the essential part of a relativistic
calculation. In this letter, we therefore explore the isospin dependence
of the spin-orbit term in non-relativistic Skyrme calculations and
analyze the consequences of this on the nuclear properties.
We start from the Skyrme type force
\begin{eqnarray}
V(1,2)&=&t_0(1+x_0P^\sigma)\delta({\bf r}_1-{\bf r}_2)\nonumber\\
&+&~t_1(1+x_1P^\sigma)(\delta({\bf r}_1-{\bf r}_2){\bf k}^2+h.c.)\nonumber\\
&+&~t_2(1+x_2P^\sigma){\bf k}\delta({\bf r}_1-{\bf r}_2){\bf k}
\label{mska}\\
&+&~\frac{1}{6}t_3(1+x_3P^\sigma)\delta({\bf r}_1-{\bf r}_2)\rho^\alpha
\nonumber\\
&+&~W_0(1+x_wP^\tau)(
{\mbox{\boldmath $\sigma$}}^{(1)}+{\mbox{\boldmath $\sigma$}}^{(2)})
{\bf k}\times\delta({\bf r}_1-{\bf r}_2){\bf k}
\nonumber
\end{eqnarray}
with ${\bf k}=\frac{1}{2}({\bf p}_1-{\bf p}_2)$. In contrast to the
conventional Skyrme ansatz, where the energy functional contains a
Hartree- and a Fock-contribution, we here neglect the exchange (Fock-) term
for the spin-orbit potential in the last line of Eq. (\ref{mska}).
Otherwise, the operator $P^\tau$ would be equivalent to $+1$ for spin
saturated systems. For the rest of the potential the Fock terms are
included. The associated 11 parameters $t_i$, $x_i$ ($i=0\dots 3$),
$W_0$, $x_w$ and $\alpha$ of the Modified Skyrme Ansatz (MSkA) in
Eq. (\ref{mska}) are determined by a fit to
experimental data of finite spherical nuclei.
The nuclear properties taken into consideration are the empirical
binding energies and charge radii of the closed-shell nuclei $^{16}$O,
$^{40}$Ca, $^{90}$Zr, and $^{208}$Pb. In order to take into account
the variation in isospin we have also included Sn-isotopes $^{116}$Sn,
$^{124}$Sn, and the doubly closed nucleus $^{132}$Sn as well as
one of the lead isotopes $^{214}$Pb. The resulting force and its
parameters are presented in Table \ref{T1}.

The spin-orbit term in the single-particle field derived from
this force has the form ${\bf W}_\tau({\bf r})(
{\bf p}\times{\mbox{\boldmath $\sigma$}})$ with
\begin{equation}
{\bf W}_\tau({\bf r})~=~
W_1{\mbox{\boldmath $\nabla$}}\rho_\tau+,
W_2{\mbox{\boldmath $\nabla$}}\rho_{\tau^\prime\ne\tau},
\label{spinorbit}
\end{equation}
where $\rho_\tau$ is the density for neutrons or protons
($\tau=n$ or $p$) and $W_1 =W_0(1+x_w)/2$, $W_2 =W_0/2$.
Conventional Skyrme calculations use a spin-orbit potential without
isospin dependence and include the Fock term. This
leads to $x_w=1$ and the relationship $W_1/W_2=2$,
which is nearly by a factor 2 different from the value
$1.0005$ obtained within the modified Skyrme Ansatz MSkA (see Table
\ref{T1}). It is interesting to note that the fit leads to a
value of $x_w$ very close to zero, which corresponds to the one
without the Fock term.

In order to study the spin-orbit term in the RMF theory we start
from the standard Lagrangian density\cite{GRT.90}
\begin{eqnarray}
{\cal L}&=&\bar\psi\left(\gamma(i\partial-g_\omega\omega
-g_\rho\vec\rho\vec\tau-eA)-m-g_\sigma\sigma
\right)\psi
\nonumber\\
&&+\frac{1}{2}(\partial\sigma)^2-U(\sigma )
-\frac{1}{4}\Omega_{\mu\nu}\Omega^{\mu\nu}
+\frac{1}{2}m^2_\omega\omega^2\nonumber\\
&&-\frac{1}{4}{\vec{\rm R}}_{\mu\nu}{\vec{\rm R}}^{\mu\nu}
+\frac{1}{2}m^2_\rho\vec\rho^{\,2}
-\frac{1}{4}{\rm F}_{\mu\nu}{\rm F}^{\mu\nu}
\label{lagrangian}
\end{eqnarray}
which contains nucleons $\psi$ with mass $m$.
$\sigma$-, $\omega$-, $\rho$-mesons, the electromagnetic field and
nonlinear self-interactions $U(\sigma)$ of the $\sigma$-field,
\begin{equation}
U(\sigma)~=~\frac{1}{2}m^2_\sigma\sigma^2+\frac{1}{3}g_2\sigma^3+
\frac{1}{4}g_3\sigma^4.
\end{equation}

In a non-relativistic approximation of the corresponding Dirac
equation\cite{Koe.94} we obtain the following single particle
spin-orbit term:
\begin{eqnarray}
{\bf W}_\tau({\bf r})&=&
\frac{1}{m^2m^{*2}}(C^2_\sigma+C^2_\omega+C^2_\rho)
{\mbox{\boldmath $\nabla$}}\rho_\tau\nonumber\\
&&~+\frac{1}{m^2m^{*2}}(C^2_\sigma+C^2_\omega-C^2_\rho)
{\mbox{\boldmath $\nabla$}}\rho_{\tau^\prime\ne\tau},
\label{relspinorbit}
\end{eqnarray}
with $C_i^2=(m\,g_i/m_i)^2$ for $i=\sigma,\omega,\rho$,
which is similar in form to the spin-orbit field (\ref{spinorbit}),
but which contains $r$-dependent parameters
\begin{eqnarray}
W_1&=&\frac{1}{m^2m^{*2}}(C^2_\sigma+C^2_\omega+C^2_\rho)\\
W_2&=&\frac{1}{m^2m^{*2}}(C^2_\sigma+C^2_\omega-C^2_\rho).
\end{eqnarray}
The $r$-dependence drops out in the ratio $W_1/W_2$ which is
$1.13$ for the parameter set NL1\cite{GRT.90}
and $1.10$ for the parameter set NL-SH\cite{SNR.93}.
This is only slightly higher than the value $1.0005$ obtained in
the Modified Skyrme Ansatz by the fit to the empirical data
(see Table \ref{T1}). The absolute size of the spin-orbit
term turns out to be $r$-dependent, which stems from the
$r$-dependence of the effective mass $m^*(r)$. It can be approximated
by $m/m^*\approx 1+C^2_\sigma/m^3\,\rho(r)$. A more careful consideration
would therefore require an explicitly density dependent spin-orbit term.
It has not been included in the present investigation.

In Table \ref{T2} we show nuclear matter results obtained in the MSkA and
compare it with the values from the conventional Skyrme force SkM$^*$.
The saturation density is obtained as $\rho_0=0.1531$ fm$^{-3}$ in MSkA.
This is the same as the value obtained from an extensive fit of the
mass formula FRDM\cite{MNM.94}. The binding energy per particle $E/A$
is 16.006 MeV, which is close to that of other Skyrme forces. The compression
modulus $K=319$ MeV is somewhat higher than that of the presently
adopted Skyrme forces, but it is in good agreement with a recent
analysis based upon the breathing mode energies\cite{Sha.91}. The third
derivative $e'''=\partial^3(E/A)/\partial \rho^3$ has a value, which
is close to that obtained in Ref. \cite{Sha.91}. The asymmetry energy
$J$ is close to the empirical value of 33 MeV. The effective mass
$m^*$ is in good agreement with that of SkM$^*$.

In Table \ref{T3} we show binding energies and charge radii
obtained in the MSkA for a number of spherical nuclei
and compare them with those from SkM$^*$.
A comparison with the empirical values shows, that the binding
energies obtained with MSkA have improved over those of SkM$^*$.
The slightly reduced charge radii in MSkA seem to be connected with
a slightly higher binding energy, which improves the results for
the lighter Pb-isotopes. In general the charge radii are improved
including that of $^{16}$O.

In Fig. 1 we show the isotope shifts of Pb nuclei for the
Modified Skyrme Ansatz MSkA together with experimental data and
results of conventional Skyrme calculations. For MSkA we observe a clear
kink at the double magic nucleus $^{208}$Pb, whereas the conventional
Skyrme force SkM$^*$ with an isospin independent spin-orbit term
gives an almost straight line. For the lighter isotopes both theories
give excellent agreement, on the heavier side MSkA comes closer to the
experimental isotope shifts.
It may be recalled, that the RMF theory\cite{SLR.93} is
successful in reproducing the full size of this kink. MSkA
uses a density independent spin-orbit force.
In contrast the spin-orbit term derived from the RMF theory
(see Eq. \ref{relspinorbit}) is implicitly density dependent
through the density dependence of the effective mass $m^*({\bf r})$.
A density dependence of the spin-orbit term in Skyrme theory might
improve the charge radii of the heavier Pb-isotopes.

In Fig. 2 we present binding energies of Zr-nuclei about the neutron
drip line as a function of the mass number. It is observed that
in agreement with earlier investigations\cite{Dob.94} within the
conventional Skyrme theory shell effects about the closed shell
nucleus $^{122}$Zr are weakened considerably. In contrast,
RMF-calculations exhibit strong shell effects in the drip
line region. It has been surmised in Ref. \cite{SLH.94} that
this is caused by the differences in the spin-orbit terms in the
two approaches. It is gratifying to see that introduction of
an isospin dependent (not antisymmetrized)
spin-orbit term in Skyrme theory leads
to stronger shell effects. This is also in agreement with
the predictions of the FRDM\cite{MNM.94}, where also an isospin
dependent spin-orbit term is used. A more quantitative analysis
shows that the ratio $W_1/W_2$ in FRDM is close to 1.06 for the
Pb-nuclei and 1.09 for $^{122}$Zr.

Summarizing, we conclude that the isospin dependence of the
spin-orbit term has an essential influence on the details of
anomalous isotope shifts of Pb-nuclei. A new Modified Skyrme
Ansatz has been proposed, and the isospin dependence of the
spin-orbit strength has been determined. The magnitude of this
isospin dependence $x_w$ is in agreement with the deductions from
the relativistic mean-field theory. A reasonably good agreement with
the experimental data on the binding energies and charge radii
has been obtained. The kink in the isotope shifts of Pb-nuclei has
been obtained in the modified Skyrme ansatz.
However, the agreement with the empirical
isotope shifts for heavy Pb-nuclei is not so good as in the
RMF theory. This calls for further investigations including
a density dependence of the spin-orbit term.

Finally a remark about the isospin dependence of the spin-orbit
term: our investigations lead to the interesting result, that
the parameter $x_w$ in the Eq. (1) is close to zero. This leads
practically to an isospin independent single-particle spin-orbit field
in Eq. (2). This is in agreement with relativistic calculations,
where the entire isospin-dependence of the spin-orbit field is
caused by the parameter $C_\rho$ in Eq. (5), which is in fact rather
small. In contrast the two-body spin-orbit potiential in conventional
Skyrme theory is isospin independent. The exchange term, however,
causes a strong isospin dependence in the corresponding
single-particle spin-orbit field. On the other hand, in the RMF theory
the spin-orbit field has its origin in Lorentz covariance. There
is no contribution from a two-body spin-orbit potential and
an exchange term is therefore excluded.

One of the authors (G.A.L) acknowledges support from the E.U., HCM
program, contract: EG/ERB CHBICT-930651 , This
work is also supported in part by the Bundesministerium
f\"ur Forschung und Technologie under the project 06 TM 733.

\figure{Isotopic shifts in the charge radii of Pb isotopes normalized
to the nucleus $^{208}$Pb as a function of the mass number A in the
Modified Skyrme Ansatz (MSkA) as compared to the Skyrme force SkM$^*$
and the empirical data.
\label{F1}}

\figure{Binding energy of Zr isotopes at neutron drip line in the
MSkA and SkM$^*$ calculation.
\label{F2}}

\begin{table}
\caption{Parameters of the interaction in the modified Skyrme
Ansatz (MSkA).}
\begin{tabular}{ll}
  $t_0 = -1200.074$ (MeV fm$^3$) & $x_0 = 0.187$ \\
  $t_1 = 396.302$ (MeV fm$^5$)   & $x_1 = 0.018$ \\
  $t_2 = -105.579$ (MeV fm$^5$)  & $x_2 = -0.059$\\
  $t_3 = 10631.527$ (MeV fm$^{3+3\alpha}$) & $x_3 =  0.046$ \\
  $W_0 = 316.38$ (MeV fm$^5$) & $x_w = 0.0005 $   \\
  $\alpha = 0.7557 $  & \\
\end{tabular}
\label{T1}
\end{table}

\begin{table}
\caption{Nuclear matter properties obtained in the modified Skyrme
Ansatz (MSkA).}
\begin{tabular}{lll}
&   MSkA & SkM$^*$\\
\hline
 $\rho_0$       & 0.1531 fm$^{-3}$ & 0.1603 fm$^{-3}$ \\
 $(E/A)_\infty$ & 16.006 MeV       & 15.776 MeV       \\
 $K$            & 319.4  MeV       & 216.7  MeV       \\
 $\rho_0^2e'''$ & 100.2 MeV        &  57.9  MeV       \\
 $J$            & 30.0 MeV         &  30.0  MeV       \\
 $m^*/m$        & 0.76             &   0.79           \\
\end{tabular}
\label{T2}
\end{table}

\begin{table}
\caption{The binding energies and charge radii obtained with the
Modified Skyrme Ansatz (MSkA) in the Hartree-Fock approximation as
compared with the normal Skyrme force SkM* and the empirical values.}
\begin{tabular}{cccccccc}
 & \multicolumn{3}{c}{Binding Energies (MeV)}&
  & \multicolumn{3}{c}{ Charge Radii (fm)} \\
\cline{2-4} \cline{6-8}
Nuclei & expt.& MSkA &  SkM* && expt. & MSkA & SkM* \\
\hline
$^{16}$O&-127.6&-128.1& -127.7  && 2.730 & 2.742 & 2.811 \\
$^{40}$Ca&-342.1&-342.8& -341.1 && 3.450 & 3.468 & 3.518 \\
$^{48}$Ca&-416.0&-416.8& -420.1 && 3.500 & 3.506 & 3.537 \\
$^{90}$Zr&-783.9&-781.9& -783.0 &&4.270 & 4.274 & 4.296  \\
$^{116}$Sn&-988.7&-984.0& -983.4 && 4.626& 4.623 & 4.619 \\
$^{124}$Sn&-1050.0&-1047.9& -1049.0 && 4.673& 4.677 & 4.678 \\
$^{132}$Sn&-1102.9&-1106.3& -1110.7 &&   -   & 4.728 & 4.727 \\
$^{200}$Pb&-1576.4&-1570.9& -1568.4 && 5.464 & 5.465 & 5.468 \\
$^{202}$Pb&-1592.2&-1588.3& -1586.0 && 5.473 & 5.475 & 5.478 \\
$^{204}$Pb&-1607.5&-1605.4& -1603.4 && 5.483 & 5.486 & 5.489 \\
$^{206}$Pb&-1622.3&-1622.3& -1620.3 && 5.492 & 5.496 & 5.501 \\
$^{208}$Pb&-1636.5&-1637.8& -1636.4 && 5.503 & 5.506 & 5.510 \\
$^{210}$Pb&-1645.6&-1645.8& -1645.6 && 5.522 & 5.522 & 5.520 \\
$^{212}$Pb&-1654.5&-1653.8& -1654.5 && 5.540 & 5.537 & 5.531 \\
$^{214}$Pb&-1663.3&-1661.6& -1663.1 && 5.558 & 5.552 & 5.541 \\
\end{tabular}
\label{T3}
\end{table}
\end{document}